# Mesoscale simulation of semiflexible chains. I. Endpoint distribution and chain dynamics

Robert D. Groot

*Unilever Research Vlaardingen, PO box 114, 3130 AC Vlaardingen, The Netherlands*

The endpoint distribution and dynamics of semiflexible fibers is studied by numerical simulation. A brief overview is given over the analytical theory of flexible and semiflexible polymers. In particular, a closed expression is given for the relaxation spectrum of wormlike chains, which determines polymer diffusion and rheology. Next a simulation model for wormlike chains with full hydrodynamic interaction is described, and relations for the bending and torsion modulus are given. Two methods are introduced to include torsion stiffness into the model. The model is validated by simulating single chains in a heat bath, and comparing the endpoint distribution of the chains with established Monte Carlo results. It is concluded that torsion stiffness leads to a slightly shorter effective persistence length for a given bending stiffness.

To further validate the simulation model, polymer diffusion is studied for fixed persistence length and varying polymer length $N$. The diffusion constant shows crossover from Rouse ($D \propto N^{-1}$) to reptation behaviour ($D \propto N^{-2}$). The terminal relaxation time obtained from the monomer displacement is consistent with the theory of wormlike chains. The probability for chain crossing has also been studied. This probability is so low that it does not influence the present results.

## I. INTRODUCTION

Fibers have a large number of applications in food and paint industry, e.g. in increasing fluid viscosity. In general fibers can increase fluid viscosity by three mechanisms. Firstly, for dilute fiber suspensions viscosity is increased by hydrodynamic interactions between the fibers. This increases viscosity by a factor that depends on the shape and volume fraction of the fibers. Both rod-like and plate-like fibrous particles lead to a higher (intrinsic) viscosity than simple spherical particles, see e.g. Groot[1] for a review.

A second mechanism that leads to viscosity increase for thin rods or semiflexible fibers is entanglements. This becomes important above the overlap concentration. A third mechanism may become pertinent in the same concentration range, if fibers associate together and form a network, i.e. when their surface energy is sufficiently large. In the latter case stress and viscosity are related to the network structure: the number of fibers across a network connection (a fiber bundle) and the mesh size, or length of network connections.

The process of network forming by fiber bundles[2] is not only important from a technological point of view, but is also important in many biological systems. For instance, plant cell walls are networks of largely crystalline cellulose fibers held together by a mixture of polysaccharides, hemicelluloses and pectins.[3,4] Another important and even more complex example is bone tissue, which is initially formed by self-assembly of collagen microfibrils and the subsequent co-crystallisation of hydroxyapatite, but ultimately shows hierarchical organisation over seven levels.[5,6]

To understand the formation of these and similar structures we need to reproduce them in simulation. To develop such a simulation model it is noted that the concentration regimes and mechanisms mentioned above are probably linked. E.g. we may hypothesise that the thickness of fiber bundles may depend on hydrodynamic interactions between the fibers, if the formation of fiber bundles is a kinetic process. In a polymer melt or dense suspension long-range hydrodynamics is screened[7,8] on a length scale of the polymer mesh size $\xi$, but in the *early stages* of network formation hydrodynamics may play an important role if the size of fiber bundle domains is kinetically determined. Hence to simulate the structure and rheology of fiber bundle networks hydrodynamics should be included.

Various methods have been proposed for the mesoscale simulation of colloids and fibers with full hydrodynamic interactions, e.g. lattice-Boltzmann[9] and particle based methods like stochastic rotation dynamics[10] (also known as multiparticle-collision dynamics, MPCD), Lowe-Anderson Dynamics (LAD),[11,12] Dissipative Particle Dynamics[12,13] (DPD) and the Fluid Particle Model[14] (FPM). The latter is a variant of DPD that includes shear friction between particles as well as radial friction. Of these methods, MPCD and LAD generally simulate fluids of high viscosity, and DPD and FPM simulate low viscosity. The method introduced by Stoyanov and Groot[12] combines LAD and DPD, and can be used for any fluid viscosity. Long range hydrodynamics between colloid particles or fibers with any of these particle-based methods depends on the fluid boundary conditions at the particle surface. An efficient method has been developed recently to impose stick boundary condition exactly at the surface of a colloidal particle,[1] which can equally be applied in any of the particle-based hydrodynamic methods mentioned. This method will be employed here.

The importance of the stick boundary condition becomes apparent for spherical particles. In general the relative viscosity of a suspension is $\eta_{suspension}/\eta_{solvent} = 1 + [\eta]\phi + \ldots$, where $\phi$ is the particle volume fraction and $[\eta]$ the intrinsic viscosity. For stick boundary conditions Einstein calculated $[\eta] = 2.5$, but for slip boundary condition Taylor arrived at $[\eta] = 1$, see e.g. Bicerano *et al.*[15] for a review. This indicates that the stick boundary condition is important for short fibers. The stick boundary condition may become less important for long polymers, but without the means to impose them this could not be investigated in simulation.

In general semiflexible polymers have two limiting cases, one where the persistence length $L_p$ is much shorter than the chain length $L$; these are flexible polymers ($L_p \ll L$). In the other limit the persistence length is much larger than the chain length ($L_p \gg L$); these are stiff rods. For *flexible polymers* the dynamics and end-point separation have been tested by Spenley[16] within the DPD model. Both in a melt and in dilute solution all exponents are in line with theory. In a melt the end-point separation $R_e$ was found to scale as $R_e \sim (N-1)^{0.498 \pm 0.005}$, where $N$ is the number of beads per polymer; the diffusion coefficient scales as $D \sim (N-1)^{-1.02 \pm 0.02}$; and the end-to-end relaxation time $\tau_R$ is proportional to $\tau_R \sim N^{1.98 \pm 0.03}$. In dilute solution $R_e \sim (N-1)^{0.58 \pm 0.04}$ and $\tau_R \sim N^{1.80 \pm 0.04}$, i.e. the expected $\tau_R \sim R_e^3$ to within the noise. This correspondence is expected as long range hydrodynamic interactions in the DPD method are known to be correct.[13] Mussawisade et al[17] simulated flexible polymer with full hydrodynamics using MPCD. This showed characteristic $t^{-3/2}$ tails in the polymer centre-of-mass velocity autocorrelation, and good correspondence with the Zimm theory.

The other limit, of rod-like polymers, has been studied in the context of DPD by AlSunaidi et al.[18] who used rigid constraints to form infinitely stiff rods, and Levine et al.[19] who simulated slightly flexible rods. Both groups obtained the known liquid-crystalline phases and observed the isotropic-nematic transition. Earlier work by Bolhuis and Frenkel[20] studied the phase boundaries and the isotropic to nematic transition of hard spherocylinders, and the hydrodynamic stress in a suspension of rods was described by Shaqfeh and Frederickson.[21] An overview of the theory and experiments in this field is given by Vroege and Lekkerkerker.[22]

Here we are interested in semiflexible fibers that are neither in the limit of flexible chains nor in the limit of stiff rods. An overview of this field was given by Shaqfeh.[23] The related Kramers chain of connected rigid cylinders was simulated with multibody hydrodynamic interactions by Butler and Shaqfeh.[24] Similar to this, an idealised model of very thin semiflexible chains was proposed by Ramanathan and Morse.[25] This latter model is based on a Monte Carlo scheme where steps are rejected if bonds cross. Such a Monte Carlo scheme however does not include hydrodynamic interactions.

This paper is the first in a series of two where the dynamics and network formation of associative semiflexible polymers is studied in simulation. In the present paper a model of semiflexible fibers is described and validated. Having established the model, it will be used the next paper to study the dynamics and stability of fiber bundle networks.[26] The simulation model is validated by comparing the polymer endpoint distribution and chain dynamics with established results. For this purpose the theory of semiflexible fiber dynamics is reviewed in section II. Next, the simulation model is presented in section III, and its results are compared to the literature results in section IV. Finally, conclusions are formulated in section V. In this paper we restrict ourselves to fibers without associative interaction.

## II. THEORY OF POLYMER DYNAMICS

This section gives a brief overview of the analytical results for the dynamics of flexible and semiflexible polymers. In its simplest form the dynamics of flexible polymers is described by the Rouse and Zimm models,[27] in which a chain is represented by $N+1$ beads with coordinates $\mathbf{r}_0(t)..\mathbf{r}_N(t)$. The chain dynamics follows the overdamped Langevin equation $\partial \mathbf{r}_n / \partial t = C \zeta^{-1} \partial^2 \mathbf{r}_n / \partial n^2 + \mathbf{f}_n(t)$, where $C$ is a spring constant between adjacent beads, $\zeta$ is the friction factor and $\mathbf{f}_n(t)$ is a Brownian noise term. This equation is conveniently solved by Fourier transformation along the chain, which renders the Eigenmodes, or normal coordinates. The Fourier transform and back transform are given by

$$\mathbf{X}_k = \frac{1}{N+1} \sum_{n=0}^{N} \mathbf{r}_n \cos(k\pi n/N) - \frac{\mathbf{r}_0 + \mathbf{r}_N}{2(N+1)}$$
$$\mathbf{r}_n = \mathbf{X}_0 + 2 \sum_{k=1}^{} \cos(k\pi n/N) \mathbf{X}_k \quad (1)$$

Each normal coordinate evolves independently and decays exponentially:

$$\langle \mathbf{X}_k(0) \cdot \mathbf{X}_k(t) \rangle = \frac{Na^2}{6\pi^2} \frac{1}{k^2} \exp(-t/\tau_k) \quad (2)$$

where $a$ is the root-mean-square distance between successive beads. The relaxation time of each mode in the Rouse model is proportional to the slowest relaxation time $\tau_R$, the rotation diffusion time or Rouse time, and is given by

$$\tau_k = \frac{\tau_R}{k^2} = \frac{\zeta a^2}{3\pi^2 k_B T} \cdot \frac{N^2}{k^2} \quad (3)$$

To characterise the dynamics of polymers the mean square displacement of beads is monitored, $R_n^2 = \langle (\mathbf{r}_n(t) - \mathbf{r}_n(0))^2 \rangle$. Substituting the dynamics of Eq (2) into the back transform in Eq (1), the model gives[27]

$$R_n^2 \approx \frac{6 k_B T}{\zeta N} t + \sqrt{\frac{12 a^2 k_B T}{\zeta \pi}} \cdot \sqrt{t} \quad (4)$$

Thus, the diffusion constant is given by $\lim_{t \to \infty} D(t) = R_n^2/6t = k_B T/\zeta N$. For small times ($t < \tau_R$) the displacement of beads increases $\propto t^{1/4}$ rather than $\propto t^{1/2}$, and the effective time-dependent diffusion constant $D(t)$ has a slow $t^{-1/2}$ time decay.

For isolated chains the Rouse model is inadequate, as it lacks hydrodynamic interaction between the beads. This is included in the Zimm model. The normal modes in the Zimm model decay as in Eq (2), but now the spectrum of relaxation times is given by

$$\tau_k = \frac{\tau_Z}{k^{3/2}} = \frac{\eta_0 (a\sqrt{N})^3}{\sqrt{3\pi} k_B T} \cdot \frac{1}{k^{3/2}} \quad (5)$$

where $\eta_0$ is the viscosity of the solvent. In this case the slowest relaxation time $\tau_1 = \tau_Z$ is the Zimm time. The mean square displacement of the beads in the Zimm model is obtained as[27]

$$R_n^2 \approx \frac{16 k_B T}{\pi \sqrt{(6\pi)} \eta_0 b \sqrt{N}} t + \frac{2}{\pi^2} \Gamma(1/3) \cdot \left(\sqrt{3\pi} k_B T/\eta_0\right)^{2/3} \cdot t^{2/3}$$
$$(6)$$

Thus, in the Zimm model the time-dependent diffusion constant is proportional to $\lim_{t \to \infty} D(t) \sim k_B T/(\eta_0 \sqrt{N})$, and it has a slow $t^{-1/3}$ time decay.

The models so far describe flexible polymers, whereas in reality many polymers and fibers in particular are semiflexible. A common model for semiflexible polymers is the wormlike chain model, introduced by Kratky and Porod.[28] The wormlike chain model describes chains of fixed length $L$ that are stiff on a length scale $L_p$. The predictions of this model compare very well to experimental light scattering results of e.g. actin fibers.[29] For a continuous semiflexible polymer the bending energy is given by

$$H = \tfrac{1}{2} K_b \int_0^L \left( d^2 \mathbf{r}/ds^2 \right)^2 ds = \tfrac{1}{2} K_b \int_0^L \left( d\mathbf{t}/ds \right)^2 ds \quad (7)$$

Here, $K_b$ is bending stiffness, $L$ is the arc length, $\mathbf{r}(s)$ describes the contour of the fiber, and $\mathbf{t}(s) = d\mathbf{r}/ds$ is a unit tangent vector along the chain. Note that this model does not include torsion rigidity, only bending rigidity. The tangent correlation along the chain follows an exponential decay

$$\langle \mathbf{t}(s) \cdot \mathbf{t}(0) \rangle = \exp(-s/L_p) \tag{8}$$

where $L_p$ is the aforementioned persistence length. In any spatial dimension $d$ the persistence length is given by[30]

$$L_p = \frac{2}{d-1} \frac{K_b}{kT} \tag{9}$$

Note that a wormlike polymer confined to a 2D surface attains twice the persistence length that it has in 3D. From Eq (8) the mean-square end-to-end distance of the wormlike chain is obtained as[31]

$$\frac{\langle R_e^2 \rangle}{L^2} = \frac{2L_p}{L}\left[1 - \frac{L_p}{L}(1-\exp(-L/L_p))\right] = g(L/L_p) \tag{10}$$

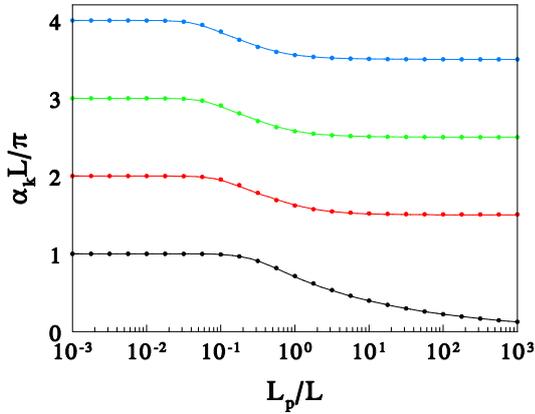

model to experimental systems. To this end we need to invert the function $g$ in Eq (10). For later reference a numerical fit is given here:

$$g^{-1}(y) \approx \frac{2}{y} - 1 - y + \frac{1}{2}\frac{y(1-y)^2}{(1-y+y^{5/6}(9.79y^2 - 14.5y^3 + 6.91y^4))} \tag{11}$$

which is accurate to four decimal places. Here, $y = \langle R_e^2 \rangle / L^2$, and $g^{-1}(y)$ is the inverse endpoint function, $g^{-1}(y) = L/L_p$.

The dynamics of the wormlike chains has been analysed by Harnau, Winkler and Reineker for the cases of dilute polymers without hydrodynamic interaction,[35] for dilute polymer solutions with hydrodynamic interaction,[36] and for polymer melts.[37] Unlike the Rouse model, the Eigenmodes are not simple cosines, but combinations of sines, cosines and hyperbolic sines and cosines. Whereas the wave numbers along the chain in the Rouse model are simply $\omega_k = k\pi/L$, for the wormlike chain they need to be solved from a set of equations, that are determined by the boundary conditions of the problem. By analysing the limiting solution for small and large persistence length and interpolating, an accurate explicit approximation to the wave numbers of the wormlike chain is obtained as

$$\frac{\alpha_k L}{\pi} \cong \begin{cases} 1/\left(1 + 30\pi^2(L_p/L)^3 + (\pi^4 L_p/24L)^{15/2}\right)^{1/30} & \text{if } k = 1 \\ k - \tfrac{1}{2} + \tfrac{1}{2}/\left(1 + 12\pi^2(kL_p/L)^3 + ((k-\tfrac{1}{2})\pi^2 L_p/4L)^6\right)^{1/6} & \text{if } k > 1 \end{cases} \tag{12}$$

The numerical wave numbers of the normal modes and the approximation given in Eq (12) are shown in Figure 1a.

If hydrodynamics is neglected the model reduces to the Rouse model in the limit $L_p/L \to 0$. For all persistence lengths the relaxation time of each Eigenmode is obtained as[35]

$$\tau_k = \frac{\zeta L_p^3}{3kT\left((L_p\alpha_k)^2 + (L_p\alpha_k)^4\right)} \tag{13}$$

where $\zeta$ is again the friction factor. The translational diffusion constant in the free draining limit is again $D = k_BT/\zeta L$, the aforementioned Rouse result. For the relaxation times, three regimes can be distinguished.[36] For $L_p\alpha_k \ll 1$ the relaxation times are proportional to $L^2/k^2$, the Rouse relaxation behaviour as in Eq (3), with $a^2$ replaced by $2L_p$. For $L_p\alpha_k \gg 1$ and $k > 1$ the relaxation times are proportional to $(L/(k-\tfrac{1}{2}))^4$, and relaxation is dominated by bending modes. In the intermediate regime both bending and stretching contribute to the relaxation times. A graph of the relaxation spectrum is shown in Figure 1b. This shows that the maximum in $\tau_k$ shifts towards lower values of the persistence length as $k$ increases, which reflects the fact that a polymer looks increasingly rigid on smaller length scales.[36] Finally, for very stiff rods ($L_p \gg L$) dynamics is dominated by a single mode, $\tau_1 = \zeta L^3/72k_BT$.

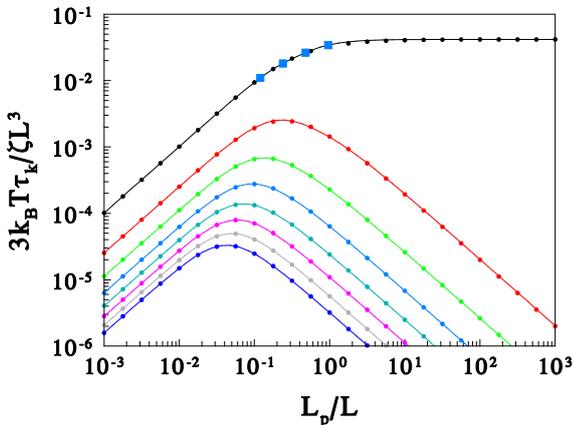

**Figure 1a** (top) Wave number of the first four normal modes as function of the persistence length. The full curves are the approximation given in Eq (12). **b** (bottom) Relaxation time of first eight modes as function of persistence length. The blue squares show the locations of the simulated systems discussed in section IV B.

By comparing the simulated end-to-end distance with the exact result of the wormlike chain, we can check if the simulated model correctly reproduces the physics of this model. Since the wormlike chain model in turn has been validated to actin,[29] DNA[32] and other systems,[33,34] this also validates the simulation

In the Rouse and Zimm model the mean square displacement as function of time shows a crossover. At early times $R_n^2 \sim t^{1/2}$ or $R_n^2 \sim t^{2/3}$ (see Eq (4) and (6)) as a consequence of internal modes, and at later times we have diffusive behaviour, $R_n^2 \sim t$. The wormlike chain model shows a similar crossover, in particular in the limit $L_p \geq L$ it leads to[37]

$$R_n^2 \approx \frac{6k_BT}{\zeta L}t + \frac{1}{(2L_p)^{1/4}\Gamma(7/4)} \cdot \left(\frac{3k_BT\,t}{\zeta}\right)^{3/4} \quad (14)$$

but for $L_p \ll L$ the chains follow Rouse dynamics (Eq (4)), with $a^2$ replaced by $2L_p$.

In a polymer melt the chains are hindered in their motion by constraints from other polymers. They can slide past but not pass through one another. This idea is captured by the tube model which was put forward by Edwards[38] in 1967. Surrounding chains restrict the transverse motion of a polymer, thus each polymer is effectively confined to a tube-like region. This model was analysed by De Gennes.[39] Because each polymer is effectively enclosed in a tube, the end-point separation vector does not change until the polymer has diffused out of its tube into a new one. This means that the polymer has to diffuse over a distance equal to its own length before the tube has completely renewed. For a polymer of $L$ beads the typical end-point relaxation time (the reptation time $\tau_{Rep}$) thus follows as $L^2 \sim D\tau_{Rep}$. Since the diffusion constant within the tube (using the Rouse model) is proportional to $1/L$, the reptation time is proportional to $\tau_{Rep} \sim L^3$. On time scales much larger than the reptation time the chain will move with an effective diffusion constant $D_{Rep}$. Since the mean square distance travelled in one reptation time is the endpoint separation of the polymer, and since the polymer in a melt behaves as an Gaussian chain, we also have $L \sim D_{Rep}\tau_{Rep}$ hence the reptation diffusion constant is proportional to $D_{Rep} \sim L^{-2}$. Though the concept was very successful in explaining experimental data, the mechanism was not actually seen until 1988. Simulation studies showed that the polymer length for flexible polymers needs to be at least 30 entanglement lengths to crossover from Rouse dynamics to reptation.[40]

It should be noted that the wormlike chain model does not contain torsion stiffness, only bending stiffness. However, just by imposing a bending stiffness and the condition that chains cannot cross a rich phenomenology is found for the rheology of dilute and semi-dilute polymers. Scaling relations for the modulus in the semi-dilute regime were given by Morse.[41,42] Several approximations can be made for the linear elastic modulus in the densely entangled regime,[43]

$$G \approx 0.40kT\rho^{7/5}L_p^{-1/5} \quad \text{or} \quad G \approx 0.82kT\rho^{4/3}L_p^{-1/3} \quad (15)$$

where $\rho$ is the total *contour length* per unit of volume.

## III. SIMULATION MODEL

### A. Particle interactions

Our starting point is the Fluid Particle Model (FPM) by Español.[14] Two types of interaction forces are used, conservative forces and dissipative forces. The dissipative forces consist of noise and friction. The friction force is implemented as a pair-wise friction between neighbouring particles, proportional to the relative velocity difference. It has been shown[44] that the friction should scale in a particular way to the particle sizes. For particles of unequal size the radial friction function is chosen as

$$\mathbf{F}_{ij}^R = -fR_{ij}\left(1 - h/R_{ij}\right)^2 \theta(1 - h/R_{ij})\,(\mathbf{v}_{ij} \cdot \mathbf{e}_{ij})\,\mathbf{e}_{ij} \quad (16)$$

where $h = r - d_{ij}$ is the gap width, $\theta(x)$ is the Heaviside step function and $\mathbf{e}_{ij} = (\mathbf{r}_i - \mathbf{r}_j)/|\mathbf{r}_i - \mathbf{r}_j|$; and we define the mean harmonic mean radius $R_{ij}$ and the mean diameter $d_{ij}$ as

$$R_{ij} = \frac{2R_iR_j}{R_i + R_j}; \quad d_{ij} = R_i + R_j \quad (17)$$

where $R_i$ is the radius of particle $i$.

The model further contains a shear friction proportional to the relative perpendicular velocities and spins. Similar models differing slightly from the original formulation by Español have been published independently by Groot and Stoyanov[45] and by Pan et al.[46] We use

$$\begin{aligned}\mathbf{F}_{ij}^S = &-\mu R_{ij}\left(1 - h/R_{ij}\right)^2 \theta(1 - h/R_{ij}) \\ &\cdot [\mathbf{v}_{ij} - (\mathbf{v}_{ij} \cdot \mathbf{e}_{ij})\,\mathbf{e}_{ij} + \mathbf{r}_{ij} \times (R_i\boldsymbol{\omega}_i + R_j\boldsymbol{\omega}_j)/d_{ij}]\end{aligned} \quad (18)$$

where $\mathbf{r}_{ij} = \mathbf{r}_i - \mathbf{r}_j$, and where $\boldsymbol{\omega}_i$ is the spin of particle $i$. The combination of velocities and spins $[\mathbf{v}_{ij} - \ldots]$ appearing in Eq (18) is the difference in particle velocity plus an extra speed given by the solid body rotation of each particle, extrapolated to the position of the other particle. $\mathbf{F}_{ij}^S$ is a force that points in a direction perpendicular to the line of contact between the particles. Brownian noise is introduced to maintain the desired temperature. In the original formulation of this model,[14,46] the noise was introduced as components of symmetric, antisymmetric and traceless Wiener matrices. This is equivalent to the following (simpler) formulation

$$\begin{aligned}\mathbf{F}_{ij}^{Ran} = &\sqrt{2R_{ij}k_BT}\left(1 - h/R_{ij}\right)\theta(1 - h/R_{ij}) \\ &\cdot \left(\sqrt{f}\,\zeta_{ij}(t)\,\mathbf{e}_{ij} + \sqrt{\mu}\,\boldsymbol{\theta}_{ij}(t) \times \mathbf{e}_{ij}\right)/\sqrt{\delta t}\end{aligned} \quad (19)$$

Here $\zeta_{ij}(t)$ is a random number of unit variance, $\boldsymbol{\theta}_{ij}(t)$ is a random vector with unit variance for each component, and $\delta t$ is the time step taken. The outer product $\boldsymbol{\theta}_{ij} \times \mathbf{e}_{ij}$ guarantees a random force perpendicular to the line of contact. Note that the distance-dependence and the amplitude follow from the fluctuation-dissipation theorem.[14,45,46] For computational efficiency $\zeta_{ij}(t)$ and each component of $\boldsymbol{\theta}_{ij}(t)$ are drawn from a uniform distribution of the correct width, the results are indistinguishable from Gaussian random numbers. Total angular momentum is conserved by a pair-wise torque, given by

$$\boldsymbol{\tau}_i = -\frac{R_i}{d_{ij}}\sum_j \mathbf{r}_{ij} \times \mathbf{F}_{ij} \quad (20)$$

Finally, we integrate the equations of motion: $\dot{\mathbf{r}}_i = \mathbf{v}_i; \dot{\mathbf{v}}_i = \mathbf{F}_i/m_i; \dot{\boldsymbol{\omega}}_i = \boldsymbol{\tau}_i/I_i$, where $m_i$ and $I_i$ are the particle mass and moment of inertia.

The conservative interaction force is found from linear elasticity theory. The repulsive force can be derived from the contact zone approximation[45] as $F^{Rep} = (3/2)Eb(r_0 - r)$, where $r_0$ is the equilibrium distance between two fused spheres, where $b$ is the radius of the neck. This means for particles that are pressed together up to centre-to-centre distance $r$ the force derivative is given by

$$\partial F^{Rep}/\partial r = -\tfrac{3}{2}Eb \quad (21)$$

The elastic repulsion for two spheres of arbitrary size and distance is thus found by integrating the force derivative from the point of first contact down to centre-to-centre distance $r$. From the geometry we find the relation $r = (R_i^2 - b^2)^{1/2} + (R_j^2 - b^2)^{1/2}$

≈ $d_{ij} - b^2/R_{ij}$, after substitution and integration we find to a lowest order approximation

$$\mathbf{F}_{ij}^{Rep} = E\sqrt{R_{ij}}(d_{ij} - r)^{3/2} \mathbf{e}_{ij} \quad (22)$$

This is the Hertz theory of elastic repulsion.[47] If $E$ is to be the elastic modulus of the material, the pre-factor should in fact be $2/(3(1-v^2))$ with $v$ Poisson's ratio. In the present formulation this factor is subsumed into the definition of $E$.

To model polymers, successive particles are connected together by harmonic springs,

$$\mathbf{F}_{ij}^{Spring} = -Cr\,\mathbf{e}_{ij} \quad (23)$$

where generally a spring constant $C = 10$ is chosen. Here we use reduced simulation units where $d = k_BT = 1$.

## B. Semiflexible fibers

To model semiflexible chains of particles *and* to maintain fluid stick boundary conditions at each particle, it is not sufficient to impose only bending interactions between two successive links. There are two fluid boundary conditions: the fluid flow velocity at the surface should vanish relative to the surface motion in radial direction and in transverse direction. If we only implement a bending interaction between the links between three successive particles, the particles are still free to rotate relative to the underlying bond structure, hence we would impose slip boundary conditions with the solvent.

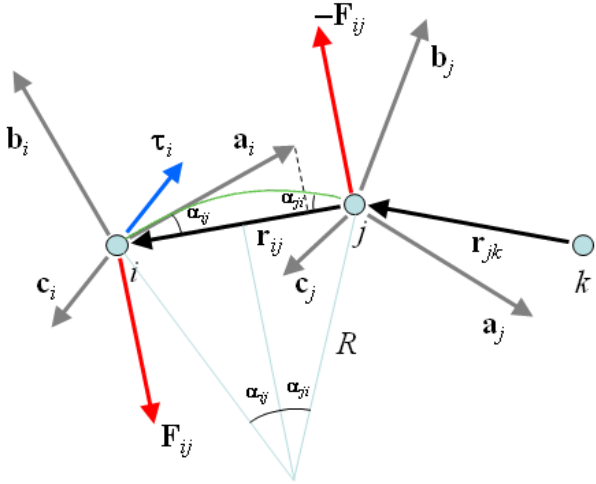

**Figure 2** Frames of particle *i* and *j* (in grey) that are connected by a stiff beam (green) after Ref [1]; the forces resulting from the misalignment of particle *i* with $\mathbf{r}_{ij}$ are shown in red, and the torque on particle *i* is shown in blue.

To restrict this rotational freedom, each particle *i* has three internal unit orientation vectors ($\mathbf{a}_i$, $\mathbf{b}_i$, $\mathbf{c}_i$). To evolve these vectors to the next time step, the particle spin $\boldsymbol{\omega}_i$ is integrated in time using a second order velocity-Verlet algorithm, together with the integration of particle positions. The frame of particle *i* that is bound to particle *j*, and the resulting forces and torque in this configuration are shown in Figure 2. The particles *i* and *j* are part of a semiflexible fiber of stiffness $K$. To maintain a fixed orientation of particle *i* relative to the chain we define the director of the fiber in point *i* by the unit vector $\mathbf{a}_i$ and impose a force to keep it aligned with the coordinate difference $\mathbf{r}_{ij} = \mathbf{r}_i - \mathbf{r}_j$. The green curve in Figure 2 denotes a stiff beam between particles *i* and *j* with local orientations $\mathbf{a}_i$ and $\mathbf{a}_j$. If this beam is fixed at points *i* and *j*, a bending moment is exerted which is given by $M = K_b/R$, where $R$ is the radius of curvature. The force exerted in point *i* is thus $F = M/r_{ij} = K_b/Rr_{ij}$. From Figure 2 we find the local curvature as $1/R = 2\sin(\alpha_{ij})/r_{ij}$, hence the total force acting on particle *i* needed to maintain the curvature is given by $F = 2K_b\sin(\alpha_{ij})/r_{ij}^2$.

The direction of this force is given by the component of $\mathbf{a}_i$ perpendicular to $\mathbf{r}_{ij}$, hence

$$\mathbf{F}_{ij}^b = -K_b\sin(\alpha_{ij})\mathbf{e}_\alpha / r_{ij}^2 = -K_b(\mathbf{a}_i - \mathbf{e}_{ij}(\mathbf{a}_i \cdot \mathbf{e}_{ij}))/r_{ij}^2 \quad (24)$$

where $\mathbf{e}_\alpha$ is a unit vector in the direction of the deflection. An opposite force $-\mathbf{F}_{ij}^b$ acts on particle *j* (see Figure 2). To conserve angular momentum, a torque acts on particle *i*, as shown in Figure 2, which tends to align the particle orientation with the chain direction:

$$\boldsymbol{\tau}_i = -\mathbf{r}_{ij} \times \mathbf{F}_{ij}^b \quad (25)$$

Note that the torque here only acts on particle *i*, since it is this particle that is aligned with the bond vector $\mathbf{r}_{ij}$.

This force accounts for half the bending moment of the fiber. A similar force and torque are defined for particle *j*, where force $\mathbf{F}_{ji}^b$ is calculated for particle *j*, and an opposite force acts on particle *i*. Thus, for each linked pair *ij* two forces and torques as defined in Eqs (24) and (25) are applied (with indices *ij* replaced by *ji*). Therefore particle *j* is forced to align with bond *ij* and also with bond *jk*. Reversely, bonds *ij* and *jk* are both forced to align with the director $\mathbf{a}_j$, which means that we effectively simulate a bending stiffness along the chain.

Whereas the Hamiltonian and the time derivative of a particle position in a Rouse chain involve a second derivative of neighboring particle positions along the chain, the stiffness of the semiflexible chain implies a fourth derivative along the chain (see Eq (7), $\partial \mathbf{r}/\partial t \sim -\delta H/\delta \mathbf{r}(s) = -K_b\partial^4\mathbf{r}/\partial s^4$. Although conceptually more insightful, such an approach requires second neighbor interactions in both directions along the chain, as $\partial^4\mathbf{r}/\partial n^4 \approx (\mathbf{r}_{n+2} - 4\mathbf{r}_{n+1} + 6\mathbf{r}_n - 4\mathbf{r}_{n-1} + \mathbf{r}_{n-2})/d^4$ if $d$ is a *fixed* distance between particle centers along the chain. It should be noted that the director $\mathbf{a}_i$ in Eq (24) corresponds to the continuum director $\mathbf{t} = \partial\mathbf{r}/\partial s$ in Eq (7). This allows the present formulation with moments that act on the paricles, which is computationally less intensive.

If $N$ particle distances along a chain are constrained to $d$, the chain has a fixed length $L = Nd$. The continuum limit is obtained as $d \to 0$, while at the same time $N = L/d$ and $K_b = L_p/d$. Thus the wormlike chain with Hamiltonian Eq (7) is recovered. In the present simulation model we allow some flexibility in the springs between successive beads so that the total arc length is not conserved. However, as the chain length is increased, the *relative* fluctuations in the arc length are proportional to $1/\sqrt{N}$, hence the wormlike chain of fixed length is recovered as a limiting case $N \to \infty$. One purpose of this paper is to check this correspondence, which is presented in Section IV A.

So far we have only discussed a bending interaction along the chain, but the particles are still free to rotate along an axis parallel to the chain. To prevent this rotation we use the other internal vectors of the particles. If two particles *i* and *j* are linked, we require the vectors $\mathbf{b}_i$ and $\mathbf{b}_j$ to align (see Figure 2). However, a misalignment where e.g. $\mathbf{b}_j$ is tilted in the $\mathbf{r}_{ij}$ direction is in fact a bending deformation. To prevent double counting of the bending forces and deal with torsion only, we first rotate the frame of particle *i* to that of *j*, to undo the bending. Thus, $\mathbf{b}_i$ is rotated around the axis $\mathbf{u} = (\mathbf{a}_i \times \mathbf{a}_j)/|\mathbf{a}_i \times \mathbf{a}_j|$ to a vector $\mathbf{b}'_i$ to undo the bending. In case $|\mathbf{a}_i \times \mathbf{a}_j| = 0$ there is no bending, hence we take $\mathbf{u} = \mathbf{b}_i$. Since parallel vector components to $\mathbf{u}$ are invariant to rotation around the $\mathbf{u}$ axis, and since ($\mathbf{a}_i$, $\mathbf{b}_i$, $\mathbf{c}_i$) forms an orthonormal frame, the outer product $\mathbf{b}'_i \times \mathbf{b}_j$ is obtained conveniently by projecting $\mathbf{b}_i$, $\mathbf{c}_i$, $\mathbf{b}_j$ and $\mathbf{c}_j$ onto $\mathbf{u}$. We thus obtain the torques and forces

$$\begin{aligned} \boldsymbol{\tau}_i &= K_t\,(\mathbf{b}'_i \times \mathbf{b}_j) \cdot \mathbf{a}_j\,(\mathbf{a}_j/r_{ij}) \\ &= K_t[(\mathbf{c}_i \cdot \mathbf{u})(\mathbf{b}_j \cdot \mathbf{u}) - (\mathbf{b}_i \cdot \mathbf{u})(\mathbf{c}_j \cdot \mathbf{u})]\,(\mathbf{a}_j/r_{ij}) \\ \boldsymbol{\tau}_j &= -K_t[(\mathbf{c}_i \cdot \mathbf{u})(\mathbf{b}_j \cdot \mathbf{u}) - (\mathbf{b}_i \cdot \mathbf{u})(\mathbf{c}_j \cdot \mathbf{u})]\,(\mathbf{a}_j/r_{ij}) \\ \mathbf{F}_i &= -\mathbf{F}_j = \mathbf{r}_{ij} \times (\boldsymbol{\tau}_i + \boldsymbol{\tau}_j)/r_{ij}^2 \end{aligned} \qquad (26)$$

where $K_t$ is the torsional stiffness. Because $\mathbf{a}_i$ and $\mathbf{a}_j$ can be misaligned by bending, the forces given in Eq (26) are needed to balance the sum of torques.

An *alternative* approach is not to define the torque on the particles along their $\mathbf{a}$-axis, but along the centre-to-centre line. In that case no balancing forces are needed, and we impose the torques

$$\begin{aligned} \boldsymbol{\tau}_i &= -\boldsymbol{\tau}_j = -K_t\,(\mathbf{b}'_i \times \mathbf{b}_j) \cdot \mathbf{a}_j\,(\mathbf{e}_{ij}/r_{ij}) \\ &= -K_t[(\mathbf{c}_i \cdot \mathbf{u})(\mathbf{b}_j \cdot \mathbf{u}) - (\mathbf{b}_i \cdot \mathbf{u})(\mathbf{c}_j \cdot \mathbf{u})]\,(\mathbf{e}_{ij}/r_{ij}) \end{aligned} \qquad (27)$$

Note that Eq (27) contains a factor $(\mathbf{e}_{ij}/r_{ij})$ where Eq (26) contains a factor $(\mathbf{a}_{ij}/r_{ij})$. The vector $\mathbf{e}_{ij} = (\mathbf{r}_i - \mathbf{r}_j)/|\mathbf{r}_i - \mathbf{r}_j|$ points from particle $j$ to particle $i$. Both Eq (26) and Eq (27) serve to produce a fiber with torsional stiffness.

In general, bending stiffness is given by $K_b = EI_a$, where $E$ is the Young's modulus of the material and $I_a$ is the area moment of inertia $I_a = \pi d^4/64$. For a solid, round fiber, we have

$$K_b = \pi E d^4 / 64 \qquad (28)$$

where $d$ is the fiber diameter. Torsional stiffness is characterised by

$$\tau = K_t \alpha / l \qquad (29)$$

where $\alpha$ is the rotation angle at the end of a clamped beam in radians, and $l$ is the length of the beam. The torsional stiffness is given by $K_t = GJ$, where $G$ is the shear modulus of the material and $J$ is the torsion constant of the sample. For round beams or fibers, $J$ is identical to the polar moment of inertia $J = \pi d^4/32$, thus for round fibers we have

$$K_t = \pi G d^4 / 32 = K_b / (1 + \nu) \qquad (30)$$

At this point we use the relation $G = \tfrac{1}{2}E/(1+\nu)$, which holds for *isotropic* materials, where $\nu$ is Poisson's ratio of the material (for most materials $\nu \sim 0.2\text{-}0.4$). Note that e.g. composite fibers may not be isotropic in which case $K_t$ may be much smaller than $K_b$.

To stabilize the spin temperature, a DPD-like thermostat force is introduced that acts only on the particles that are connected by a bond with torsional stiffness. No distance-dependence is used. Thus we add to the torques in Eqs (26) and (27) a random torque, given by

$$\delta\boldsymbol{\tau}_i = -\delta\boldsymbol{\tau}_j = -f_t\left[(\boldsymbol{\omega}_i - \boldsymbol{\omega}_j) \cdot \mathbf{e}_{ij}\right]\mathbf{e}_{ij} + \sqrt{2 f_t kT}\,\zeta_{ij}(t)\,\mathbf{e}_{ij}/\sqrt{\delta t} \qquad (31)$$

where $f_t$ is a torsion factor, and $\zeta_{ij}(t)$ is a random number of unit variance. This randomizes the spin parallel to the centre-to-centre axis of two neighbouring particles.

## C. Long range hydrodynamics

To include full hydrodynamics, ideal gas solvent particles are added that interact with the colloidal particles through a hard core interaction. However, to obtain full hydrodynamic interaction between the colloid particles, the radial and transverse relative velocity of the fluid should vanish on average at the particle surface.[1]

To impose the correct *radial* boundary condition, the fluid particles are simply reflected from the particle surface. To implement this into a DPD or FPM model is non-trivial; the correct implementation depends upon the order of position, force and velocity update. The correct order is given in Ref[1]. The *transverse* boundary condition requires a generalisation of the Lowe-Anderson thermostat[11] to transverse velocities.[1] When a fluid particle collides with a colloid particle, a Monte Carlo step is taken to exchange momentum and replace the transverse velocity difference by a new value taken from a Gaussian distribution. Thus, two interacting particles $i$ and $j$ are given new velocities according to

$$\begin{cases} \mathbf{v}_i' = \mathbf{v}_i + \Delta\mathbf{p}/m_i \\ \mathbf{v}_j' = \mathbf{v}_j - \Delta\mathbf{p}/m_j \end{cases} \qquad (32)$$

Let the colloid particle be $i$ and the fluid particle be $j$. The fluid particle is assumed to behave as ideal particle. Because it has zero size, its spin does not change by collision with a colloid particle. Since the force on the colloid particle $i$ acts on its surface, its angular momentum changes by $-\mathbf{r}_{ij} \times \Delta\mathbf{p}$, thus its spin changes according to

$$\boldsymbol{\omega}_i' = \boldsymbol{\omega}_i - \mathbf{r}_{ij} \times \Delta\mathbf{p}/I_i \qquad (33)$$

It was shown by Groot[1] that the correct velocity distribution is generated when the (transverse) momentum exchange $\Delta\mathbf{p}$ is taken as

$$\Delta\mathbf{p} = -\mu\left[\mathbf{v}_{ij} - \mathbf{e}_{ij}(\mathbf{e}_{ij} \cdot \mathbf{v}_{ij}) + \mathbf{r}_{ij} \times \boldsymbol{\omega}_i + \boldsymbol{\theta}_{ij} \times \mathbf{e}_{ij}\sqrt{kT/\mu}\right] \qquad (34)$$

where each component of the vector $\boldsymbol{\theta}_{ij}$ is a standard normally distributed random variable and $\mathbf{e}_{ij} = (\mathbf{r}_i - \mathbf{r}_j)/r_{ij}$ and $\mathbf{v}_{ij} = \mathbf{v}_i - \mathbf{v}_j$. The effective mass $\mu$ is defined as

$$1/\mu = 1/m_i + 1/m_j + r_{ij}^2/I_i \qquad (35)$$

where again $m_i$ and $m_j$ are particle masses, and $I_i$ is the moment of inertia of the colloid particle.

At high colloid volume fraction the effective viscosity is dominated by the direct (viscous) interaction between the colloid particles, because the fluid particles are relatively sparse. Here, viscosity is primarily determined by the range and amplitude of the viscous interaction between the colloid particles, whereas at low volume fraction the indirect interaction via fluid particles dominates, and viscosity depends on the fluid particle density. Thus, at one value of the fluid particle density the low volume fraction viscosity and the high volume fraction viscosity are self-consistent, matching up the viscosity at low and high colloid concentrations.[1] This happens at solvent density $\rho_s r_c^3 \approx 2.2$. Since the cut-off range for viscous interaction is taken as $r_c = 1.5$, the solvent particle density is chosen as $\rho_s = 0.652(1-\phi)$, where $\phi$ is the colloid volume fraction.

## IV. SIMULATION RESULTS

### A. Isolated chains

To check the present simulation model, single chains were simulated in a cubic box and the mean square end-point separation was averaged. We used chains of $N = 10, 20, 30$ and 40 beads of diameter $d = 1$. In each case the repulsion parameters between beads was chosen as $E = 2000$, and neighbouring particles on the chains were bound together by harmonic spring forces $F_{ij} = Cr_{ij}$ with $C = 10$. The chains of 10 and 20 beads were enclosed in boxes of size $V = 15 \times 15 \times 15$, the chain of 30 beads was simulated in a box of size $V = 20 \times 20 \times 20$, and the 40 bead chain was simulated in a box of size $V = 30 \times 30 \times 30$. In all cases periodic boundary conditions were applied. Some 70 to 100 ideal gas particles were added to improve temperature control. Temperature was fixed at $k_BT = 1$ and the time step used was $\delta t = 0.01$ unless stated otherwise. Sensitivity to the time step was analysed previously.[1] The stick boundary conditions of the fluid at the particle surface were found to be exactly satisfied up to $\delta t = 0.025$, and colloid viscosity obtained from $\delta t = 0.01$ and $\delta t = 0.005$ are identical. All results are given in simulation units where the colloid particle diameter $d$ is used as unit of length, and the unit of time is $t^* = d\,(m/k_BT)^{1/2}$, where $k_BT = 1$ is the set temperature, see Eq (19).

The root mean square bond length $a$ varied slightly between the systems; this was measured in the same run to obtain the total chain length $L = (N-1)a$. Using Eqs (10) and (11), the persistence length can be obtained from $L/L_p = g^{-1}(R_e^2/L^2)$. In all four cases we used the bending stiffness $K_b = 10$ and the torsional stiffness $K_t = 0$. This should correspond to the wormlike chain model. The results obtained over $15 \times 10^6$ time steps are summarised in Table 1. All results are consistent with $L_p = K_b/k_BT = 10$.

| $N-1$ | $a$ | $L$ | $R_e$ | $L_p$ |
|---|---|---|---|---|
| 9 | 1.0335±0.0005 | 9.301±0.001 | 8.06±0.02 | 10.0±0.2 |
| 19 | 1.0365±0.0005 | 19.693±0.002 | 14.81±0.06 | 9.8±0.2 |
| 29 | 1.0363±0.0005 | 30.052±0.003 | 20.22±0.14 | 9.9±0.2 |
| 39 | 1.0331±0.0005 | 40.289±0.004 | 24.87±0.14 | 10.2±0.2 |

**Table 1** Simulation output for four single chain systems and the resulting persistence lengths for bending stiffness $K_b = 10$. $N$ is the number of beads per chain, $a$ is the root-mean-square bond length and $R_e$ is the root-mean-square endpoint separation.

As a second test, the shape of the endpoint distribution $P(r)$ was compared with the known result for wormlike chains. Note that in 3D the probability to find a separation distance $r$ is given by $S(r) = 4\pi r^2 P(r)$. $P(r)$ is related to the potential of mean force by $\Psi(r) = -k_BT \ln[P(r)]$. Dhar and Chaudhuri[48] determined the endpoint separation $P(r)$ for chains of 1000 segments with fixed angles between neighbouring bonds using Monte Carlo simulation. They evaluated $10^8$ conformations. Their results for $L/L_p = 1$, 2 and 3.33 are reproduced here in Figure 3, and compared to the distributions found for the present simulation model, averaged over $15 \times 10^6$ time steps. The two models match quantitatively. In these simulations we used chains of $N = 20$ and $N = 33$ beads with $K_b = 10$ to obtain $L/L_p = 2$ and 3.33 and $N = 20$ with $K_b = 19.69$ to obtain $L/L_p = 1$, and we added 80, 70 and 80 ideal gas particles for better temperature control. These runs took some 20 min on a single E5345, 2.33 GHz processor.

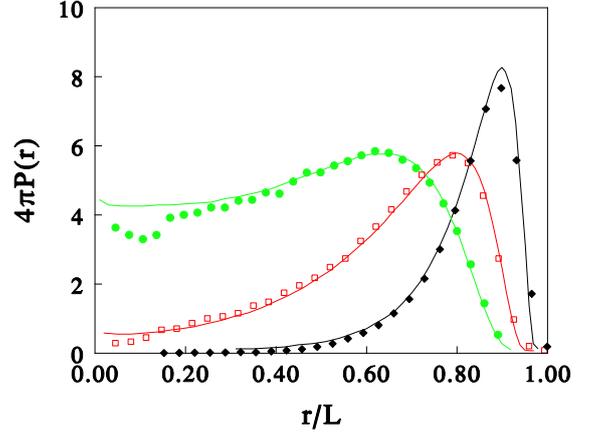

**Figure 3** Endpoint distribution for chains with $L/L_p = 1$ (black), 2 (red) and 3.33 (green) for present simulation model (symbols) and for ideal wormlike chains (lines) as obtained from FIG 2 of Ref[48].

It should be noted that the natural unit of length of a semiflexible polymer is its persistence length, see e.g. Eq (13). This means that in any physical system we should compare the polymer length with the persistence length. In these simulations the unit of length is taken as the colloid particle diameter. Taking $K_b = 10$ thus means that we fit 10 particles in one persistence length. Within the context of these simulations, taking a larger bending constant (which was done for $L/L_p = 1$ in Figure 3, where $K_b = 19.69$) thus means that we simulate to a higher resolution. Simulating at higher value of $K_b$ in these simulations therefore does not probe a truly different system, but one that is closer to the continuum limit.

Now that the validity of the model with respect to the endpoint separation has been established we can investigate the influence of torsional stiffness. Since rotation of a set of segments around their own axes is irrelevant for the final endpoint separation, we expect the same persistence length with and without torsion stiffness. Over three runs of 10, 20 and 30 bead chains respectively, for $K_t = K_b = 10$ we find persistence length estimates of $L_p = 9.54±0.04$; $L_p = 9.56±0.08$ and $L_p = 9.55±0.07$ respectively. Thus, we find that a finite torsion stiffness reduces the persistence length. The effect is small, however, even at $K_t = K_b = 10$ the reduction in $L_p$ is some 4.5±0.3%. This may be related to the finite extensibility of the chains. However, when the endpoint distributions with and without torsion stiffness are compared, for the same ratio of chain length to persistence length, a quantitative match is obtained. This means that adding torsional stiffness only rescales the persistence length, but the endpoint distribution itself is still given by the wormlike chain model.

The above results were obtained *without* applying the spin thermostat of Eq (31). To test if the slight decrease in the above persistence lengths can be explained by a lack in temperature stability the solvent was removed altogether and a test was run over $10^5$ steps of size $\delta t = 0.01$. The system contained 128 polymers of length 20 (2560 particles), but no solvent. Just having a bending interaction ($K_b = 10$, no torsion, no solvent) *increases* the spin temperature $k_BT_s$ by 2.5% and *decreases* the kinetic temperature (measured from the linear velocities) by 3.9%. Thus we have a moderate temperature error, probably caused by "spin-orbit coupling" induced by the bending forces. For an ideal gas of the same density, the spin temperature remains close to 1 ($k_BT_{spin} = 1.009$ and $k_BT = 1.007$), hence bending interaction leads to some difficulties.

| System | Torsion Eq (26) | | Torsion Eq (27) | |
|---|---|---|---|---|
| | $k_B T_{kin}$ | $k_B T_{spin}$ | $k_B T_{kin}$ | $k_B T_{spin}$ |
| solvent | 1.007 | 1.009 | 1.007 | 1.009 |
| bending only | 0.961 | 1.025 | 0.961 | 1.025 |
| bending&torsion | 1.008 | 1.181 | 0.994 | 1.247 |
| b&t+spin therm. | 1.008 | 1.082 | 0.977 | 1.051 |
| id.+ solvent (1) | 1.007 | 1.040 | 0.993 | 1.039 |
| id.+ solvent (2) | 1.0052 | 1.0081 | 0.9995 | 1.0079 |

**Table 2** Kinetic temperature and spin temperature for various systems, where the torsion interaction is introduced via Eq (26) or Eq (27). The first line (solvent) gives results for solvent only; the next three lines are for polymers only with bending interaction, bending and torsion, and bending, torsion and the spin thermostat of Eq (31). The last two lines give the temperature, averaged over the polymer particles only, in a polymer/solvent system with 2440 solvent particles (system 1) and with 16679 solvent particles (system 2).

When a torsion interaction $K_t = 7.5$ is introduced using Eq (26) (without the spin thermostat Eq (31); maintaining $K_b = 10$), spin temperature increases to $k_B T_{spin} = 1.181$ while kinetic temperature is back to $k_B T = 1.008$. When torsion is introduced using Eq (27) we obtain $k_B T_{spin} = 1.247$, while kinetic temperature is $k_B T = 0.994$, see Table 2. Whereas bending interaction decreases the kinetic temperature, the torsion interaction increases it again. However, whereas the errors in the kinetic temperature are of the order 1%, the errors in the spin temperature are of the order 20%. Therefore the spin thermostat Eq (31) was introduced.

When the spin thermostat is applied for a moderate value of the spin friction coefficient $f_t = 1$, the temperature error is largely reduced, see Table 2. Using Eq (26) the kinetic temperature remains unchanged upon applying the spin thermostat (0.8% error), but the error in spin temperature decreases from 18% to 8%. With Eq (27) the error in the kinetic temperature is increased and becomes 2% to low, but the error in spin temperature decreases from 25% to 5%. When an extra 2440 solvent particles are added the errors in the mean kinetic and spin temperatures − averaged over the polymers only − decrease by a factor 2. In Table 2 these results are marked by solvent (1). In most practical cases, the solvent density will be chosen as $\rho_s = 0.65$ to represent correct hydrodynamics between the polymers.[1] This implies for the present system 16679 solvent particles; in Table 2 this is marked by solvent (2). In this latter case the error in the temperature is below 1% for both the spin temperature and the kinetic temperature, irrespective of the equation used to implement the torsion forces. This is quite acceptable. It should moreover be noted that for thermodynamic functions like the bond-length distribution and end-point distribution only the kinetic temperature is relevant.

When the endpoint distribution is simulated again with the spin thermostat turned on, identical results are obtained, i.e. a persistence length $L_p \approx 9.55$ whereas $L_p = 10$ was expected. Hence, a lack of temperature control cannot be a source of problems. Thus is seems that a torsion stiffness equal to the bending stiffness $K_t = K_b$ reduces the persistence length by some 4.5%. The effect is small, but well outside the error in temperature control.

## B. Entangled fiber diffusion

The dynamics of network formation, as well as the viscosity of (entangled) polymer solutions, depend on the diffusion constant of the polymers in the network formed by all other polymers. Thus, to understand the timescale of network formation and to predict viscous behaviour we first need to analyse the diffusion behaviour of entangled polymers. To this end, the mean square displacement $\delta R$ of each particle was followed in time. The obtained value $D(t) = \delta R^2/6t$ was averaged over successive time intervals, where each interval was doubled in time as compared to the previous.

Four systems were studied. Each system comprised 2560 colloidal particles and 16679 fluid particles to generate long range hydrodynamic interactions in a box of volume V = 30×30×30 ($\phi = 0.05$). The colloidal particles in the same polymer chain were bound together by elastic springs to form semiflexible chains with persistence length $L_p \approx 9.7$ ($K_b = 10$, $K_t = 7.5$). In these simulations the elasticity parameter of the beads is taken as $E = 1000$, which results into a root mean square bond length $a = 1.012$, and spin friction coefficient $f_t = 0.2$ was used. Although all systems contain 2560 colloidal particles they differ in chain length and number of chains. System 1 comprises 256 chains of $N = 10$ beads; system 2 comprises 128 chains of $N = 20$ beads; and systems in 3 and 4, $N = 40$ and $N = 80$ are simulated respectively. To gain simulation speed the time step in these simulations was taken as $\delta t = 0.025$, which is still a safe value, leading to temperature $k_B T_{kin} = 1.0023 \pm 0.0005$ and $k_B T_{spin} = 1.0099 \pm 0.0007$. The systems were evolved over 10 485 760 (= $20 \times 2^{19}$) time steps, and monomer positions were sampled every 20 steps, stored for every power of 2 samples, and the mean square displacement for each logarithmically chosen time interval was averaged over the run. The time-dependent diffusion constants are shown in Figure 4. The displacement shown is the mean square displacement of the (colloid) monomers, averaged over all monomers in the system. It should be noted that the displacement is inversely proportional to the solvent viscosity. For the present system this is given by[1] $\eta \approx 0.315 \pm 0.005$.

In general we expect the mean square displacement to increase as $\delta R^2 = 6D(t)t = 6D_\infty t + 6\lambda t^n$, where the exponent $n$ equals $n = 1/2$ for Rouse dynamics, $n = 2/3$ for Zimm dynamics, and $n = 3/4$ for semiflexible polymers of $L_p > L$, see Eqs (4), (6) and (14). The long time diffusion constant $D_\infty$ can be read off the data in Figure 4b where $\delta R^2/6t$ runs horizontally. To obtain this limiting diffusion constant the data was fitted logarithmically to a *convenient* fit function $D(t) = D_\infty + \lambda t^\alpha \exp(-t/\tau)$, which interpolates between the early stage power law decay and long time diffusion. A measure for the terminal relaxation time $\tau_R$ is obtained by fitting the data to a power law and intersecting this with the previously obtained value for the long time diffusion constant. The fit results are given in Table 3. The power $n$ given in the third column of Table 3 *qualitatively* shows crossover from $L_p > L$ behaviour ($n = 3/4$) towards Rouse behaviour ($n = 1/2$). In fact the observed exponents are 0.14 higher than expected ($n \approx 0.89 \pm 0.01$ for $N = 10$ instead of 3/4, and $n \approx 0.63 \pm 0.01$ for $N = 80$ instead of 1/2). This may be attributed to the fact that for short polymers contributions from internal dynamics and center-of-mass dynamics add up, which leads to a stronger time dependence, i.e. higher exponents.[17]

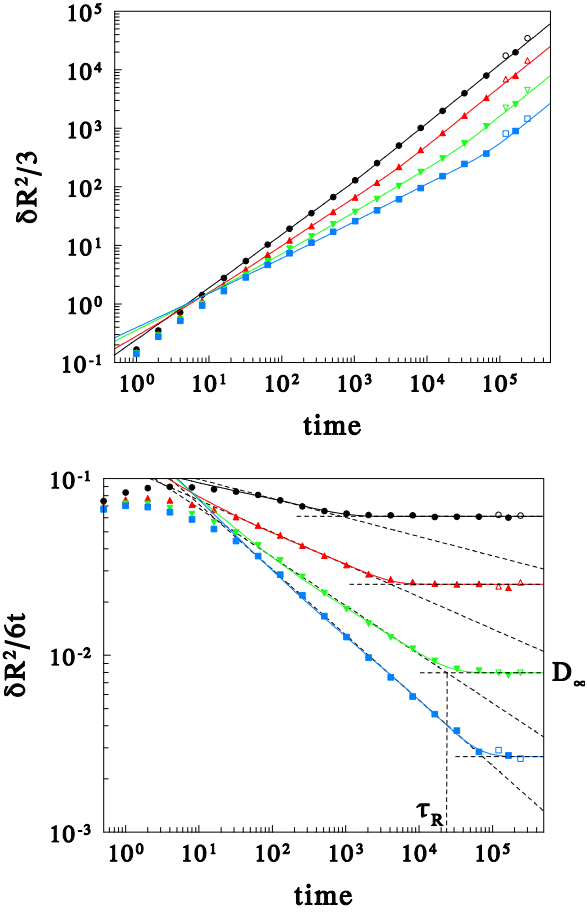

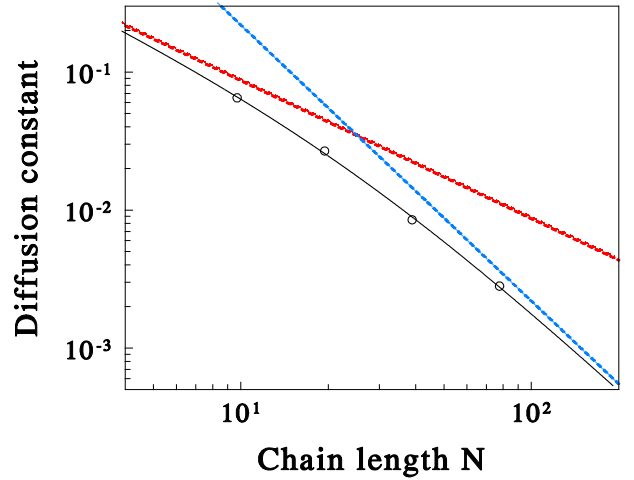

| N | $D_\infty$ | n | $\tau_R$ | $R_e$ | $\phi^*$ |
|---|---|---|---|---|---|
| 10 | 0.0612(3) | 0.89(1) | $8.5(1.5) \cdot 10^2$ | 7.88(2) | 0.013 |
| 20 | 0.0252(2) | 0.82(1) | $40(5) \cdot 10^2$ | 14.42(6) | 0.0043 |
| 40 | 0.0080(2) | 0.72(1) | $2.5(3) \cdot 10^4$ | 23.8(1) | 0.002 |
| 80 | 0.0026(2) | 0.63(1) | $8(2) \cdot 10^4$ | 37.5(1) | 0.001 |

**Table 3** Fit parameters for entangled polymer systems of bending stiffness $K_b \approx 9.7$. $N$ is the number of beads per chain, $D_\infty$ is determined by fitting the terminal diffusion behaviour; $n-1$ is the slope of the dashed curves in Figure 4b and $\tau_R$ is estimated from the intersection of these curves with $D_\infty$. $R_e$ and $\phi^*$ are the estimated endpoint separation and overlap concentration.

**Figure 4** Distance travelled (top) and time dependent diffusion constant (right) for chains of $N = 10$ beads (bottom), $N = 20$ (red), $N = 40$ (green), and $N = 80$ (blue) with $L_p \approx 9.7$. The open symbols for $t > 10^5$ are actual simulated results, the closed symbols in between are the average of these.

**Figure 5** Diffusion constant for chains of $L_p \approx 9.7$, for chains of $N = 10, 20, 40$, and 80 beads. Red dashed line is limiting slope $-1$; blue dashed line is slope $-2$. The fit curve shows crossover from Rouse to reptation diffusion.

At finite volume fraction, entanglements could play a role in diffusion. The critical volume fraction where chains start to overlap follows from the volume fraction of enveloping spheres of the chains. When this equals the dense random packing volume fraction $\phi_{envelop} = \phi(R_e/d)^3/N = 0.64$ the chains start to overlap, hence we estimate the overlap concentration as

$$\phi^* \approx 0.64 N(d/R_e)^3 \qquad (36)$$

Thus, for $N = 10$ we estimate $\phi^* = 0.013$. This implies that the volume fraction $\phi = 0.05$ used here is 4 times above $\phi^*$ for $N = 10$ and some 50 times above $\phi^*$ for $N = 80$. Comparing this to previous work, Kremer et al.[40] simulated flexible chains at volume fraction $\phi \approx 0.45$, and found crossover to reptation at chain length $N_e = 35$ and end-point separation $R_e \approx 7.7$. For this chain Eq (36) predicts $\phi^* = 0.05$, hence in the study of Kremer et al. crossover occurred at $\phi_c \approx 9\phi^*$.

The diffusion constant for the present system is shown in Figure 5. The black curve running through the data is $D = D_0/[N(1+N/N_e)]$, with $N_e = 25\pm2$, the chain length at crossover. The dashed red and blue curves give the respective short and long chain length limits. The chain length $N_e = 25$ corresponds to an overlap concentration $\phi^* = 0.003$. Thus, at crossover the simulated volume fraction is $\phi_c \approx 16\phi^*$, which compares reasonably with the value $\phi_c \approx 9\phi^*$ that was obtained for flexible chains. The ratio $\phi_c/\phi^*$ may well depend on persistence length.

The relaxation time $\tau_R$ obtained by intersecting the power law behaviour of $D(t)$ with limiting value $D_\infty$ is shown in Figure 6. The black curve is a fit to Eq (13), substituting the value for $\alpha_1(L/L_p)$ from Eq (12). The only fit parameter is the vertical scale of the curves. This shows a reasonable correspondence between theory and simulation. More in general we expect $\tau_R \sim L^3$ for $L < L_p$; then a crossover to $\tau_R \sim L^2$ for $L > L_p$; and finally a crossover to again $\tau_R \sim L^3$ for $L \gg L_e > L_p$. It should be remarked that the uncertainty in the last point ($N = 80$) may be higher than indicated, due to low statistics on the diffusion constant for this system (only 32 chains are simulated). Moreover, the diffusion constant for this system may be overestimated as we see only a small stretch of horizontal slope in Figure 4b. A better estimate for this system may be obtained by tracking the mean square displacement of the polymer centre of mass.

An alternative way of analysing the observed relaxation time is to assume crossover from Rouse dynamics to reptation,

$$\tau_R \sim N^2(1+N/N_e) \qquad (37)$$

where $N_e = 25\pm2$ is the chain length at crossover, determined above from the diffusion behaviour. This relation shows an excellent fit to the first three data points, and is shown in Figure 6 by the red dashed curve, but now the last point is off by a factor 2. Hence, whereas the diffusivity starts to show reptation behaviour, the terminal relaxation time does not. This apparent inconsistency may be caused by the fact that the systems are chosen in the crossover region from rod-like polymers to flexible polymers. To illustrate this, the positions of the simulated systems are indicated in Figure 1b. The consequence of this is that the downward slope in Figure 4b gets steeper as the polymer length increases. Therefore the crossing point with the limiting value of $\delta R^2/6t$ does not increase as much with polymer length as would otherwise be the case if all systems had the same slope. Hence the apparent inconsistency may be related to the definition of the terminal relaxation time used here. Another difference between theory and simulation is inertia, which is absent in an overdamped Langevin model. The data shows fast dynamics at very early stages, $\delta R^2 \sim t^n$, with $n$ even above 1 for $t < 1$ (see Figure 4b). This 'super diffusive' behaviour is most likely a remnant of polymer inertia, and appears to affect shorter polymers more than longer polymers.

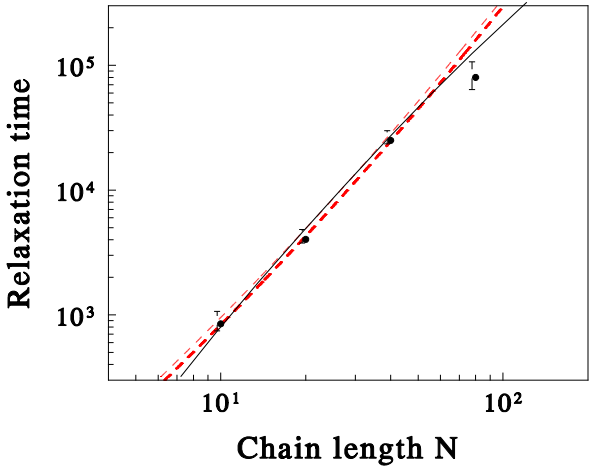

**Figure 6** Relaxation time for chains of bending stiffness $K_b = 10$ and torsional stiffness $K_t = 7.5$, for chains of $N = 10$, 20, 40, and 80 beads. Dots are obtained by intersecting the power law behaviour of $D(t)$ with limiting value $D_\infty$. The black curve is a fit to Eq (13) using persistence length $L_p \approx 9.7$, the red dashed curve is a fit to Eq (37).

## C. Fiber crossing probability

At this point one may wonder if diffusivity has been increased significantly by artificial crossings of chains. In principle the springs binding the particles together (Eq (23)) can be extended, so that the bonds between two pairs of neighbouring particles can pass through each other. It is possible to eliminate such chain crossings completely,[25,49,50] but this has not been implemented here. To estimate the time for this to occur by thermal fluctuations, we need to calculate the collision time and the probability for two bonds to cross per collision.

First we estimate the collision time. This is the time for a polymer to diffuse over a distance equal to the correlation length $\xi$. Within the length scale of the mesh size $\xi$, the polymer concentration is $c = N_\xi/\xi^3 = (\xi/d)^{1/\nu}/\xi^3$, where $N_\xi$ is the number of monomer units, $d$ is the monomer size and $\nu$ is the swelling exponent. Hence, for rodlike polymers the mesh size (and hydrodynamic screening length) is given by $\xi = d^{1/(1-3\nu)}c^{\nu/(1-3\nu)} = (dc)^{-1/2}$. For strings of particles of diameter $d$ we thus obtain $\xi^2 = (dc)^{-1} = \pi d^2/(6\phi) \approx 10.5d^2$ for $\phi = 0.05$. Hence the typical time between uncorrelated collisions $t_\xi$ is solved from $\xi^2 = \delta R^2/3 = 10.5d^2$; the chains have then moved over a distance $\xi$ in *any* direction. Fitting the data in Figure 4a to power laws we find for $N = 10$, $t_\xi = 65$; for $N = 20$, $t_\xi = 106$; for $N = 40$, $t_\xi = 165$; and for $N = 80$, $t_\xi = 230$. The displacement data shows fast dynamics at very early stages, $\delta R^2 \sim t^n$, with $n$ slightly above 1 for $t < 1$. This 'super diffusive' behaviour is most likely a remnant of polymer inertia, related to the low fluid viscosity. The actual time that a polymer moves over a correlation length will thus be an intricate function of polymer stiffness, length and viscosity.

To calculate the probability of two chains crossing we consider two pairs of particles as shown in Figure 7. Each pair is separated by a distance $r_1$ and $r_2$ respectively, which follow the probability $P(r_1,r_2) \sim \exp(-H/k_BT)$, where

$$H = \sum_i u(r_i) + \tfrac{1}{2}Cr_i^2 = \sum_i \tfrac{\sqrt{2}}{5}E(1-r_i)^{5/2}\theta(1-r_i) + \tfrac{1}{2}Cr_i^2 \quad (38)$$

where $u(r_i)$ is the repulsive interaction potential between the particles. The probability of crossing is the probability that particle pair 1,2 can be inserted in between pair 3,4. There is a penalty for the overlaps of particle 1 with 3 and 4, but also for particle 2 with 3 and 4. Thus, we can replace the simultaneous insertion problem of two particles, by the insertion of a single particle between four others, where particles 4 and 5 are images of particles 2 and 3, shifted over $\mathbf{r}_2$. The crossing probability thus takes the form of a Widom insertion, given by

$$P_{cross} = \frac{\int \exp(-U/kT - H/kT)\,dr_1\,dr_2\,d\theta\,dx\,dy}{\int \exp(-H/kT)\,dr_1\,dr_2\,d\theta\,dx\,dy} \quad (39)$$
$$= \langle \exp(-U/kT) \rangle_H$$

where $(x,y)$ runs over the shaded parallelogram in Figure 7 and $U$ is the overlap energy, $U = u(r_{12})+u(r_{13})+u(r_{14})+u(r_{15})$.

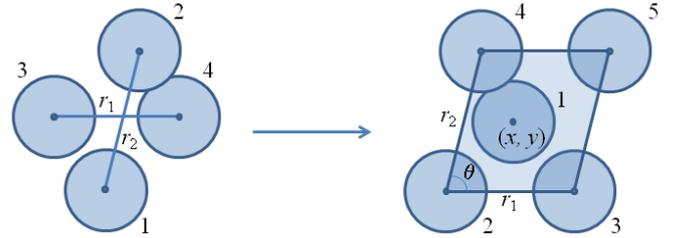

**Figure 7** Conformation of two pairs of particles is equivalent to one particle interacting with four others.

The average in Eq (39) was determined by a straightforward Monte Carlo sampling over $10^8$ conformations generated for Hamiltonian $H$. To generate the conformations, $r_1$ and $r_2$ were updated by random steps, with a step size that was determined by requiring 50% acceptation. The full run was split in 20 blocks, and average and spread were determined over these blocks to estimate the error. The results are shown in Figure 8. On this scale the error bars are smaller than the symbols. The fit curve is given by $P_{cross} = \exp(-3.23C^{3/5} - 0.027C^{9/5})$. For $C = 10$ (the value used in the simulations) the fit predicts $P_{cross} = 4.7 \cdot 10^{-7}$. The direct Monte Carlo result for $C = 10$ for $10^9$ conformations is $P_{cross} = (5.91\pm0.07)\cdot 10^{-7}$, but here the result is largely dominated by rare events so the error is not very reliable.

The picture emerging is that the crossing probability is of the order of $5 \cdot 10^{-7}$. By studying the angle distribution of successful insertions it is found that the angle $\theta$ is peaked around $90°\pm5°$. If two polymers collide once there may be a series of successive collisions, but since the polymers are locally stiff rods, all these collisions will have very similar bond angles. Thus, if at the first collision the rods are not perpendicular, the chains will have to diffuse over a correlation length ξ before a new statistically independent collision can occur. Therefore the typical time between collisions that lead to artificial crossing is $t_\xi/P_{cross}$. The number of collisions per chain in this time is roughly $Nd/\xi$, hence the time between *any* artificial crossing per chain is $\tau_c = \xi t_\xi/NdP_{cross} = (\pi/6\phi)^{1/2} t_\xi/NP_{cross}$.

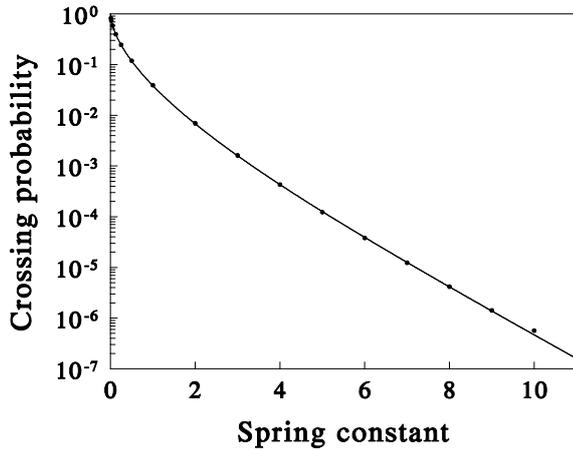

**Figure 8** Probability per collision of two chains crossing as function of the spring constant *C*.

For the presently simulated systems we find $\tau_c = 4.2 \cdot 10^7$, $3.4 \cdot 10^7$, $2.7 \cdot 10^7$ and $1.9 \cdot 10^7$ for $N = 10$, 20, 40 and 80 respectively. The relaxation times of these systems were obtained as $\tau_R = 850$, $4 \cdot 10^3$, $2.5 \cdot 10^4$ and $8 \cdot 10^4$ respectively. As the estimated crossing time is much larger than the relaxation time, this shows that artificial chain crossings do not contribute to the observed polymer diffusion. So when can we expect the present model with harmonic springs of $C = 10$ to fail? To make an estimate we need to extrapolate the collision time to long polymers. Assuming this to become independent of chain length for large *N*, and extrapolating the data by an exponential function gives $t_\xi \approx 280$ for $N \to \infty$. Thus, we estimate $\tau_c \approx 2 \cdot 10^9/N$ for large *N*. From the fit to Eq (37) the terminal relaxation time can be estimated as $\tau_R \approx 6N^2(1+N/N_e)$. This $\tau_R$ will exceed the time between artificial crossings $\tau_c$ for chain length above $N \approx 290$. Hence in that case a model should be imposed that prevents chain crossings. Even then, these will only occur in runs 30 times longer than reported here.

## V. DISCUSSION AND CONCLUSIONS

Viscosity increase by fibrous materials depends, amongst other factors, on the hydrodynamic interactions between the fibers. For this aim a method has been developed recently to simulate semiflexible fibers including hydrodynamic interactions. A first step is taken here to validate the simulation model to existing theory and experiments.

First an overview is given over the analytical theory of flexible and semiflexible polymers. In particular, a closed expression is given for the relaxation spectrum of wormlike chains, which in turn determines diffusion behaviour and polymer rheology. It is noted that several authors have shown that the wormlike chain model captures the structure and dynamics of real systems like DNA, actin and collagen fibers.

Next the simulation model for wormlike chains is described, and relations for the bending and torsion modulus are given. Two methods are introduced to include torsion stiffness into the model. The model is validated by simulating single chains in a heat bath, and by comparing the endpoint distribution of the chains with established Monte Carlo results. It is concluded that torsion stiffness leads to a slightly shorter effective persistence length for a given bending stiffness, but chains with and without torsion stiffness follow the same endpoint distribution.

Next, polymer diffusion is studied for a 5% (v/v) fiber suspension of persistence length $L_p \approx 9.7$ and polymer length $N = 10$ to 80. The diffusion constant shows crossover intermediate between Rouse behaviour ($D \propto N^{-1}$) and reptation behaviour ($D \propto N^{-2}$). To observe actual reptation behaviour at this polymer concentration would require longer polymers. The terminal relaxation time $\tau_R$ is consistent with Rouse behaviour ($\tau_R \propto N^2$); reptation behaviour for long chains ($\tau_R \propto N^3$) has not been observed. This may be related to the way the relaxation time is determined, and to the fact that the systems are chosen in the crossover regime from stiff rods to flexible chains. Also, for short polymers inertia might play a role at low fluid viscosity.

The model used here does allow some chain crossings. Therefore the crossing rate has been determined by Monte Carlo sampling. The crossing probability is so low that this does not influence the present results. Only for chain length of order $N \sim 290$ these artificial crossings would become important for the present simulation parameters.

One may wonder if the present model can be generalised to good, bad and θ solvent conditions, but this is not straightforward. At present ideal gas particles are used as solvent, which assures they do not influence the conformation of the polymers. Moreover, a hard sphere repulsion with stick boundary conditions is used between colloid and fluid, which allows definition of the exact location of the fluid boundary condition. To move away from the present θ solvent would at least require repulsion between the solvent particles. This may induce depletion and other structural effects, which would require a careful analysis even to define the θ solvent.